\documentclass[]{mycls}
\usepackage{amsmath,amssymb,amsfonts,graphicx,mathbbol,natbib,rotating,subfigure}
\usepackage{bbold}
\graphicspath{{FIGURES/}}
\usepackage{color}

\def\br{{\bf r}}
\def\bR{{\bf R}}

\def\b0{{\bf 0}}
\def\bP{{\bf P}}
\def\o2{{1 \over 2}}

\newcommand{\beq}{\begin{eqnarray}}
\newcommand{\eeq}{\end{eqnarray}}

\def\EPSFIG[#1]#2#3#4
{
  \begin{figure}
    \includegraphics[#1]{#2}
    \caption{#3}
    \label{#4}
  \end{figure}
}

\begin{document}

\markboth{E. Liberatore et al.}{Free energies in CEIMC}

\title{{\itshape Free energy methods in Coupled Electron Ion Monte Carlo}}
\author{Elisa Liberatore$^{\dagger}$, Miguel A. Morales $^{\ddagger}$, David M. Ceperley$^{\S}$, Carlo Pierleoni$^{\P}$$^{\ast}$\thanks{$^\ast$Corresponding author. Email:carlo.pierleoni@aquila.infn.it
\vspace{6pt}} \\
\vspace{6pt}$^{\dagger}$ Department of Physics, University of Rome ``La Sapienza'',  I-00185 Rome, Italy and \\ CASPUR,  Via dei Tizii, 6 - 00185 Rome, Italy\\
\vspace{6pt}$^{\ddagger}$ Lawrence Livermore National Laboratory, Livermore, California 94550, USA \\
\vspace{6pt}$^{\S}$ Department of Physics and NCSA, University of Illinois at Urbana-Champaign, Urbana, IL 61801, USA \\
\vspace{6pt}$^{\P}$ CNISM and Department of Physics, University of L'Aquila, Via Vetoio 10, I-67010 L'Aquila, Italy\\
}
\maketitle

\begin{abstract}
Recent progress in simulation methodologies and in computer
power allow first principle simulations of condensed systems
with Born-Oppenheimer electronic energies obtained by Quantum
Monte Carlo methods. Computing free energies and therefore
getting a quantitative determination of phase diagrams is one
step more demanding in terms of computer resources. In this
paper we derive a general relation to compute the free energy
of an ab-initio model with Reptation Quantum Monte Carlo (RQMC)
energies from the knowledge of the free energy of the same
ab-initio model in which the electronic energies are computed
by the less demanding but less accurate Variational Monte Carlo
(VMC) method. Moreover we devise a procedure to correct
transition lines based on the use of the new relation. In order
to illustrate the procedure, we consider the liquid-liquid
phase transition in hydrogen, a first order transition between
a lower pressure, molecular and insulating phase and a higher
pressure, partially dissociated and conducting phase. We
provide new results along the $T=600K$ isotherm across the
phase transition and find good agreement between the transition
pressure and specific volumes at coexistence for the model with
RQMC accuracy between the prediction of our procedure and the
values that can be directly inferred from the observed plateau
in the pressure-volume curve along the isotherm. This work
paves the way for future use of VMC in first principle
simulations of high pressure hydrogen, an essential
simplification when considering larger system sizes or quantum
proton effects by Path Integral Monte Carlo
methods.\end{abstract}

\bigskip

\section{Introduction}
\label{sec:introduction} 
First-principles simulation has become
an essential method to investigate the physical behavior of
condensed matter systems, in particular when the chemical
nature of the interaction among the nuclear degrees of freedom
depends on the external conditions, like for instance in
systems under extreme conditions of pressure and temperature.
Those methods generally treat systems of classical nuclei at
finite temperature and electrons in their instantaneous ground
state in the framework of the Born-Oppenheimer (BO)
approximation, although extensions to treat quantum nuclei at
finite temperature within the Path Integral
formalism, and/or electron at finite temperature with fractional (fermi) occupation have been developed. 
The electronic structure at fixed nuclear positions
is generally computed within the Density Functional Theory
(DFT) framework \cite{Martin} or, when higher accuracy is
required, by Quantum Monte Carlo methods (QMC) \cite{foulkes}.
DFT is, in principle, an exact theory but, in practice, it is
based on uncontrolled approximations for the
exchange-correlation functional. Despite some well documented
limitations, DFT is often accurate and fast enough to be used
in conjunction with Molecular Dynamics to perform dynamical
studies for systems of several hundred  nuclei and extract
physical information (first-principles Molecular Dynamics,
FPMD) \cite{VuilleumierLNP96,Kohanoff2006,MarxHutter}. Ground
state  QMC comes in two different flavors: Variational Monte
Carlo (VMC) and Projection Monte Carlo (either Diffusion Monte
Carlo (DMC) or Reptation Quantum Monte Carlo (RQMC)). VMC is
faster but the accuracy of its predictions is limited by the
skill in designing and optimizing suitable many-body trial wave
functions. RQMC is a method to automatically optimize trial
wave functions; it provides the most accurate description of
the electronic properties but has larger computational
requirements (about an order of magnitude larger than VMC, not
counting the VMC optimization process). It is an exact method
for bosons since it is able to provide the properties of the
bosonic ground state in a time which is polynomial in the
number of particles. In the case of fermions (electrons) the
scaling is exponential and for practical calculations we have
to resort to the "fixed node approximation". This approximation
makes the method variational with respect to the trial nodes;
usually the results remain highly accurate if good trial nodal
surfaces are employed \cite{foulkes}.  With recent progress in
optimization procedures \cite{UmrigarPRL} the location of the
trial nodal surfaces, suitably parametrized, could also be
optimized to achieve higher accuracy.

In recent years we have developed an ab-initio method in which
electronic energies are computed by ground state Quantum Monte
Carlo (QMC) methods while nuclear degrees of freedoms are
sampled by Metropolis Monte Carlo. This method, called Coupled
Electron-Ion Monte Carlo (CEIMC) \cite{PierleoniLNP2006}, has
been successfully applied to high pressure hydrogen. The
results are a benchmark of DFT calculations in the region of
phase diagram of interest for planetary physics
\cite{PierleoniPRL04,MoralesPRE2010} and across the
metal-insulator transition region in the fluid phase
\cite{DelaneyPRL2006,MoralesPNAS2010}.

The general picture emerging is that FPMD for high pressure
hydrogen is generally accurate away from the metal-insulator
transition while QMC accuracy is required when approaching the
metallization of the system.

In CEIMC we sample the equilibrium distribution for the
protonic degrees of freedom by a Metropolis based Monte Carlo
scheme employing either VMC or RQMC to compute the difference
between the BO energies of two nearby protonic configurations
in the Metropolis method. The electronic calculation needs to
be repeated for each attempted move of the protons and it
represents by far the largest part of the computational load of
the method. It is therefore very tempting to exploit VMC rather
than RQMC in CEIMC. On the other hand if the trial wave
function is not accurate enough, the use of VMC could provide
biased results which can lead to inaccurate predictions of the
physical behavior of the system \cite{DelaneyPRL2006}.

The key quantity in tracing transition lines is the free energy
difference (either Helmholtz or Gibbs) between coexisting
phases which is usually obtained by computing the absolute free
energies of the coexisting phases. This can be achieved by the
Coupling Constant Integration (CCI) method, a general strategy
introduced by Kirkwood \cite{Kirkwood} to adiabatically
transform the system into another system of known free energy.
This strategy is often used in ab-initio simulations to compute
the free energy of the ab-initio model starting from the
knowledge of the free energy of an effective classical
representation of the same system, for which standard free
energy methods can be easily applied
\cite{FrenkelSmit,Alfe,RedmerPRB2010,Morales2009,MoralesPNAS2010}.

In this paper, within the framework of the CCI and of the CEIMC
method, we derive a relation which allows us to obtain the free
energy of the system with RQMC electronic energies from the
knowledge of the free energy of the system with VMC energies.
The simple relation, derived in the next section, is of
practical relevance when tracing transition lines since it
allows us to obtain the transition lines with RQMC accuracy by
performing free energy calculations with VMC electronic
energies. This strategy improves the efficiency of the method
by roughly $10-20$ times, depending on the accuracy of the
trial wave function.

To illustrate the use of the new relation we consider the
liquid-liquid phase transition in high pressure hydrogen,
predicted earlier \cite{LandauZeldovich,Ebeling,SaumonChabrier}
using the so-called chemical model, but only recently put on
firm basis by first-principle simulations
\cite{MoralesPNAS2010,RedmerPRB2010}. This is a first order
transition between an insulating, molecular liquid phase at
lower pressure and a metallic, partially dissociated liquid
phase at higher pressure. It is predicted to occur below 2000K;
on increasing pressure the specific volume is discontinuous,
corresponding to a partial dissociation of molecules and to an
abrupt jump of roughly four orders of magnitude of the DC
conductivity \cite{MoralesPNAS2010}. Above T=2000K the process
is observed to become continuous and a critical point
is predicted to be located at $T_c=2000K$ and $P_c(DFT)=85GPa$,
$P_C(QMC)=120GPa$. The transition line T(P) has a negative
slope, meaning that lowering the temperature requires higher
pressure to cross the transition, and it is predicted to meet
the melting line of the molecular insulating crystalline
structure (Phase I) at a triple point located at $T_t
=700K$ and $P_t=220GPa$ within the DFT framework and $T_t=550K$
and $P_t=290GPa$ using CEIMC \cite{MoralesPNAS2010}. The latter
prediction was based on an extrapolation to high pressure of
lower pressure results.

In this paper we performed new CEIMC calculations of high
pressure hydrogen for T=600K and across the liquid-liquid phase
transition. We performed  calculations using both VMC and RQMC
with the aim of testing the accuracy of VMC across this
particular phase transition. Having at our disposal the
equation of state obtained by the two methods we can test the
reliability of our new relation for correcting free energies.

The paper is organized as follows. In the next section 
we will set the formal framework of RQMC and we will derive the
formula to correct the free energy. In section
\ref{sec:results}, as an application of our strategy, we will
present recent results obtained for the liquid-liquid
transition line of high pressure hydrogen. Finally in section
\ref{sec:conclusions}, we will draw some conclusions.

\section{Theoretical framework}\label{sec:theory}
Let us consider a system of $N$ monovalent ions of mass $M$ and
$N$ electrons of mass $m$ in a volume $V$ and at thermal
equilibrium with a bath at temperature $T$. If we assume the
validity of the Born-Oppenheimer approximation and consider
electrons in their ground state, the electrons provide a
many-body potential for the ionic motion. The ionic
``Hamiltonian" is \beq
\mathcal{H}_i(\{\bP_i,\bR_i\})=\sum_{i=1}^N \frac{\bP_i}{2
M}+\frac{1}{2}\sum_{i\neq j} \frac{e^2}{|\bR_i-\bR_j|} +
E(\{\bR_i\}) \label{eq:Hionic} \eeq where $\{\bP_i\}$ and
$\{\bR_i\}$ are the set of momentum and coordinate operators of
the ions, respectively, and $E(\{\bR_i\})$ is the electronic
energy which depends parametrically on the nuclear positions.
The thermodynamics of the system is derived from the knowledge
of the Helmholtz free energy $F(T,\rho)$ \beq F(T,\rho)=-k_B T
\log\left[ Tr\left( e^{-\beta \mathcal{H}_i}\right)\right]
\label{eq:FreeEnergy} \eeq where $\beta=1/k_BT$ and the trace
is over nuclear degrees of freedom. The term $E(\{\bR_i\})$ can
be computed at various levels of approximation, for instance
using empirical effective potentials or Density Function Theory
(DFT) or Quantum Monte Carlo (QMC) methods. Within the latter
class of methods we can distinguish between Variational Monte
Carlo (VMC) and Reptation Quantum Monte Carlo (RQMC) which
provide different level of accuracy.

In VMC we make an ansatz for the many-body wave function
$\psi_T(\br^N)=\langle\br^N|\psi_T\rangle$ and compute the
electronic energy as a statistical average of the local energy
$E_{loc}(\br^N)=[\mathcal{H}_e\psi_T(\br^N)]/\psi_T(\br^N)$
over the distribution $|\psi_T(\br^N)|^2/\int|\psi_T(\br^N)|^2$
\beq E_{T}
=\frac{\langle\psi_{T}|\mathcal{H}_e|\psi_{T}\rangle}{\langle\psi_{T}|
\psi_{T}\rangle}=\frac{\int d\br^N
\psi_{T}^*(\br^N)\mathcal{H}_e\psi_{T}(\br^N)}{\int|\psi_T(\br^N)|^2
d\br^N}= \frac{\int d\bR |\psi_{T}(\br^N)|^2
E_{loc}(\br^N)}{\int|\psi_T(\br^N)|^2 d\br^N} \eeq where
$\br^N=(\br_1,\cdots,\br_N)$ is the vector of the 3N-electronic
positions and $\mathcal{H}_e$ indicates the electronic
hamiltonian
\begin{equation}
\mathcal{H}_e=-\frac{\hbar^2}{2m}\sum_{i=1}^N \nabla_i^2+\frac{1}{2}\sum_{i\neq j} \frac{e^2}{|\br_i-\br_j|}-\sum_{i=1}^N\sum_{\alpha=1}^N\frac{e^2}{|\br_i-\bR_{\alpha}|}
\end{equation}
In an electron-ion system, although not indicated explicitly, the trial wave function 
depends parametrically on the ionic coordinates and generally
on parameters that need to be optimized by the use of the
variational theorem.

In RQMC the trial wave function is automatically optimized by
the application of the projection operator
$e^{-t\mathcal{H}_e}$ which filters out its excited states
components.
All excited states will be suppressed exponentially fast with
increasing $t$, the rate of convergence to the ground state
depending on the energy gap between the ground and the first
excited state non-orthogonal to the trial function. Here t is a
parameter with the units of inverse energy; we will refer to it
as the projection time because of the analogy between the
projection operator $e^{-t\mathcal{H}_{e}}$ and the propagator
of the real time dynamics of the system,
$e^{-it\mathcal{H}_{e}/\hbar}$. The total energy function of
time is defined as
\begin{equation}
E(t)=\frac{\langle\psi(t/2)|\mathcal{H}_e|\psi(t/2)\rangle}{\langle\psi(t/2)|\psi(t/2)\rangle}=
\frac{\langle\psi_{T}|e^{-\frac{t}{2}\mathcal{H}_e}\mathcal{H}_e e^{-\frac{t}{2}\mathcal{H}_e}|\psi_{T}\rangle}
{\langle\psi_{T}|e^{-t\mathcal{H}_e}|\psi_{T}\rangle}=
\frac{\langle\psi_{T}|\mathcal{H}_e e^{-t\mathcal{H}_e}|\psi_{T}\rangle}
{\langle\psi_{T}|e^{-t\mathcal{H}_e}|\psi_{T}\rangle}\geq E_0
\label{eq:E(t)}
\end{equation}
where $E_0$ is the exact ground state
\footnote{In case of fermions one must employ the fixed node approximation to avoid the sign problem and $E_0$ is the fixed node ground state energy which is an upper bound of the true ground state energy \cite{foulkes}}.

Similar to a thermal partition function, let us define the generating
function of the moments of $\mathcal{H}_e$ at time $t$ as
\begin{equation}
Z(t)=\langle\psi_{T}|e^{-t\mathcal{H}_e}|\psi_{T}\rangle.
\end{equation}
The total energy at time $t$ is simply the derivative of the logarithm of $Z(t)$
\begin{equation}
E(t)=-\frac{\partial}{\partial t}ln Z(t)
\end{equation}
and the variance of the energy is the second derivative
\begin{equation}
\sigma^{2}_{E}(t)= \langle(\mathcal{H}_e-E(t))^{2}\rangle= -\frac{\partial}{\partial t}E(t) \geq 0
\label{eq:sigma2}
\end{equation}
which is non-negative by definition and implies that the
energy decreases monotonically with time. The (fixed node) ground state is
reached at large time (much larger than the inverse gap) and
\begin{eqnarray}
\lim_{t\rightarrow\infty} E(t)&=&E_{0}\\
\lim_{t\rightarrow\infty} \sigma^{2}_E(t)&=&0
\label{eq:ZVRQMC}
\end{eqnarray}
The last relation is the generalization of the zero variance
principle to RQMC. Note that the average in
eq.(\ref{eq:sigma2}) is over the electronic ground state, not
to be confused with the trace over ionic degrees of freedom in
eq. (\ref{eq:FreeEnergy}).

The VMC level of description is obtained at $t=0$. This
observation allows us to derive an expression to connect the
free energy of the system with VMC electronic energy to the
free energy of the system with RQMC electronic energy. Let us
call $\mathcal{H}_i(t)$ the nuclear hamiltonian of eq.
(\ref{eq:Hionic}) with electronic energy $E(\{\bR_i\},t)$
obtained after a time $t$ projection, and $F_t(T,\rho)$ the
corresponding free energy. The time derivative of the free
energy is \beq \frac{\partial F_t}{\partial t}=\overline{\left(
\frac{\partial E(\{\bR_i\},t)}{\partial t} \right)_t}=-
\overline{\left( \sigma^{2}_{E}(t)\right)_t} \eeq where $\overline{\left( \cdots \right)}$
 indicates the trace over ionic degrees of freedom
with statistical weight $e^{-\beta\mathcal{H}_i(t)}$. The free
energy at any positive time t is obtained from the knowledge of
the VMC free energy $F_0(T,\rho)$ as \beq
F_t(T,\rho)=F_0(T,\rho) - \int_0^t ds \overline{(
\sigma^{2}_{E}(s))_s} \label{eq:rqmcFE} \eeq where the
second term on the r.h.s. is the integral of the average, over
ionic sampling, of the variance of the electronic energy. Since
the variance is positive by definition, the time derivative of
the free energy is negative, implying that the free energy is a
monotonically decreasing function of the projection time $t$:
$F_t<F_0, \forall t>0$, that is the free energy of the VMC
system is an upper bound to the fixed node free energy which is
obtained at infinite projection time. 

Eq. (\ref{eq:rqmcFE}) is the main result of our work. It is
particularly useful to increase the level of accuracy of
transition lines from VMC to RQMC. Indeed once a transition
point has been detected with the VMC energies, RQMC simulations
at fixed thermodynamic points inside the two competing phases
can be run for increasing projection time and the corrections
to the free energy of both phases can be computed as we will
show in the next section. This correction will determine how
the transition point moves in going from VMC to RQMC accuracy.
This method is much faster than performing thermodynamic
integration directly within RQMC.

Note that, being a generalization of the CCI method, eq. (\ref{eq:rqmcFE}) 
holds only if the system does not cross a phase boundary during the time projection.

\section{Results for high pressure hydrogen}\label{sec:results}
To test the use of eq. \eqref{eq:rqmcFE}, we consider the first
order liquid-liquid phase transition in high pressure hydrogen.
We have extended previous CEIMC calculations
\cite{MoralesPNAS2010} to the $T=600K$ isotherm, in the range
of pressures $P\in[220,330]GPa$ corresponding to a range of
specific volumes $v\in[9.42,11.49]a_{0}^{3}$ and of
$r_{s}=(3v/4\pi)^{1/3}\in[1.31,1.40]$ where we expect the
transition to be located. We performed CEIMC simulations in the
NVT ensemble, on systems of $54$ classical protons and $54$
electrons. We adopted backflow-Slater-Jastrow electronic trial
wave functions \cite{Holzmann2003,PierleoniDelaney}, with
analytical RPA and numerically optimized two-body correlations
and backflow terms. The single particle orbitals of the Slater
determinants were obtained by self-consistent  DFT-LDA
calculations. To reduce the finite size effects we adopted
Twist-Averaged Boundary Conditions \cite{Lin} with a fixed grid
of $64$ twist angles. Moreover, finite size corrections were
added to energy and pressure \cite{Chiesa,Drummond}. More
details can be found in ref. \cite{MoralesPNAS2010}.

\begin{figure}
\center
\includegraphics[width=0.49\textwidth]{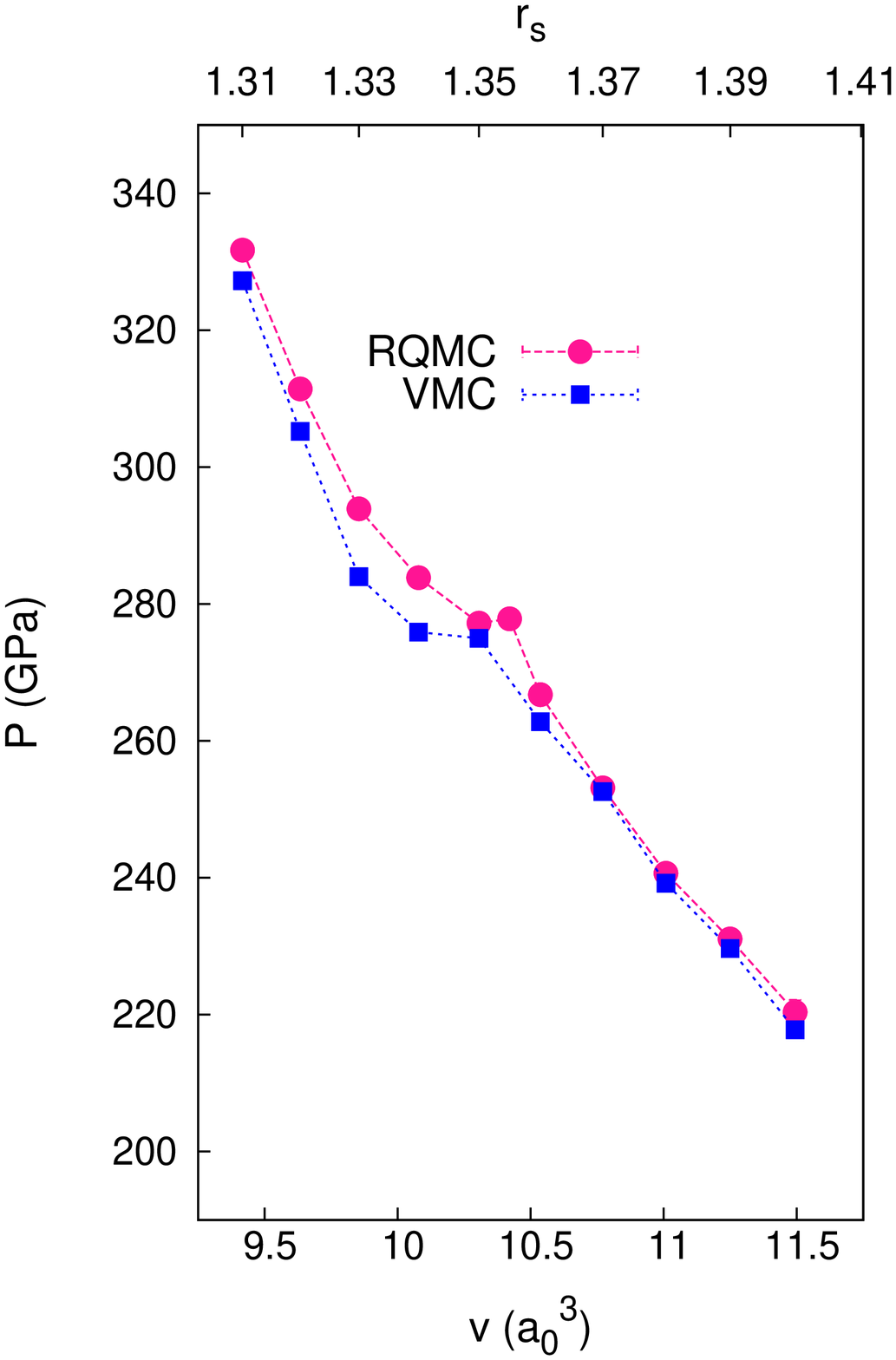}
\includegraphics[width=0.49\textwidth]{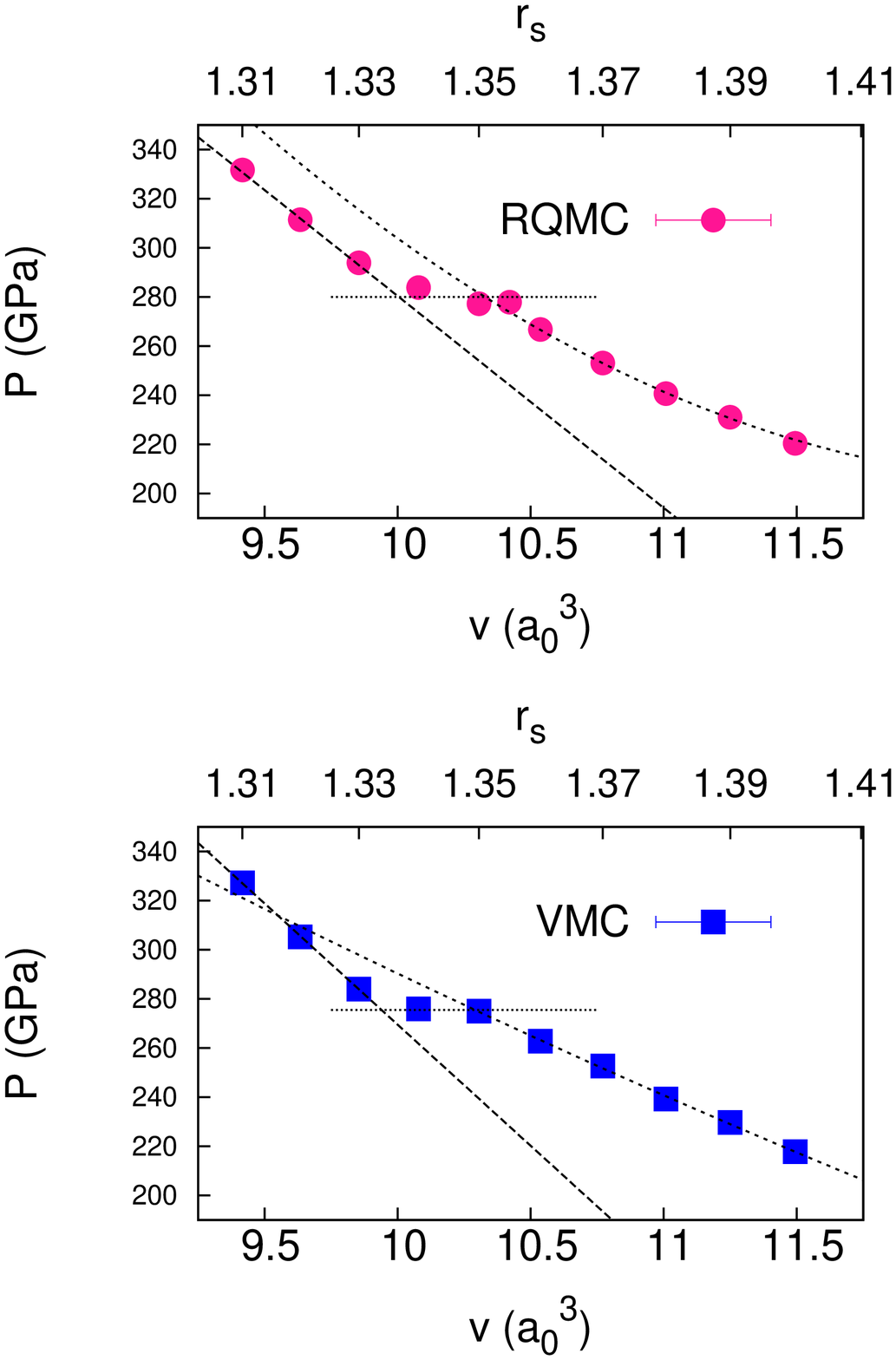}
\caption{Pressure $P$ versus specific volume $v$, along the
isotherm $T = 600K$ for VMC and RQMC data. Statistical error
bars are smaller than the symbol size. Left panel: RQMC (pink
circles) vs VMC  (blue squares) comparison. Lines are guides to
the eyes. Right panels: polynomial fits (dashed black lines) of
the equations of state for RQMC (top panel) and VMC (bottom
panel) data. The estimated coexistence pressure is represented
by the horizontal dotted line,  corresponding to $P=275.5GPa$
for the VMC curve and $P=280GPa$ for the RQMC curve. The
volumes of the two phases at coexistence are estimated as the
intersection of this line with the polynomial fits and are
$v_{II}^{c}\sim10.00(4)a_{0}^{3}$ and
$v_{I}^{c}\sim10.31(4)a_{0}^{3}$ for RQMC data and
$v_{II}^{c}\sim9.94(2)a_{0}^{3}$ and
$v_{I}^{c}\sim10.30(2)a_{0}^{3}$ for VMC data.}
\label{fig:prcfr}
\end{figure}

On the left panel of Fig. \ref{fig:prcfr} we show the pressure
versus the specific volume along the $T=600K$ isotherm, as
obtained by CEIMC with VMC energies (blue squares) and RQMC
energies (pink circles). In the RQMC calculations the total
projection time $t$ ($t=1.00H^{-1}$) and the time step $\tau$
($\tau=0.02H^{-1}$) were chosen to provide converged energies,
while pressures  were corrected for finite $t$ and nonzero
$\tau$. Numerical values are summarized in Table
\ref{tab:PPTdata}. The two curves have the same qualitative
behavior and are very close especially at large specific
volume, where pressures from the two methods agree within
statistical errors. Both curves have a plateau at
$P\sim280GPa$, signifying a weakly first order phase transition
(divergence of the isothermal compressibility). A peculiarity
of this transition, probably related
to its weakly first order character, is the absence of
metastable states which typically appear in other first order
transitions, such as, for instance, the gas-liquid transition, and makes 
difficult to obtain a quantitative location of
the transition point without computing the free energies of the
competing phases. It follows that the coexistence pressure and
the specific volume of the coexisting phases can be
qualitatively located directly from the $P(v)$ behavior without
resorting to the free energy methods.

The values of the coexistence pressure, inferred from the
position of the plateau in the $P(v)$ curves in Figure
\ref{fig:prcfr}, are of about $P^{c}_{(\text{RQMC})}\sim
280GPa$ and $P^{c}_{(\text{VMC})}\sim275.5GPa$  for the RQMC
and the VMC data respectively. The EOS for the two different
phases can be fitted with polynomial functions. At pressures
below the transition, the system is in its normal phase (phase
I), molecular and insulating. The curves $P(v)$ can be well
represented by a second order polynomial  $P_{I}(v)=p_{I}+q_{I}
v+r_{I}v^{2}$, with $p_{I}=2604GPa$,
$q_{I}=-382.5GPa/a_{0}^{3}$ and $r_{I}=15.2GPa/a_{0}^{6}$ for
RQMC data and $p_{I}=1008GPa$, $q_{I}=-91.9GPa/a_{0}^{3}$ and
$r_{I}=2.01GPa/a_{0}^{6}$ for VMC data. For the phase II, the
conducting fluid, a first order polynomial
$P_{II}(v)=p_{II}+q_{II} v$ is sufficient to describe the
pressure behavior; the values of the parameters are
$p_{II}=1143GPa$ and $q_{II}=-86.2GPa/a_{0}^{3}$ for RQMC data
and  $p_{II}=1256GPa$ and $q_{II}=-98.6GPa/a_{0}^{3}$ for VMC
data.  The specific volumes of the two phases at coexistence
can be estimated by extrapolating those fits up to the
coexistence pressure,  as shown in the right panel of Figure
\ref{fig:prcfr}.  The volume change at transition is in both
methods $\sim 4\%$. The  specific volumes of the phases at
coexistence are  $v_{II(\text{RQMC})}^{c}\sim10.00(4)a_{0}^{3}$
and $v_{I(\text{RQMC})}^{c}\sim10.31(4)a_{0}^{3}$ for RQMC data
and $v_{II(\text{VMC})}^{c}\sim9.94(2)a_{0}^{3}$ and
$v_{I(\text{VMC})}^{c}\sim10.30(2)a_{0}^{3}$ for VMC data.

\begin{figure}
\center
\includegraphics[width=\textwidth]{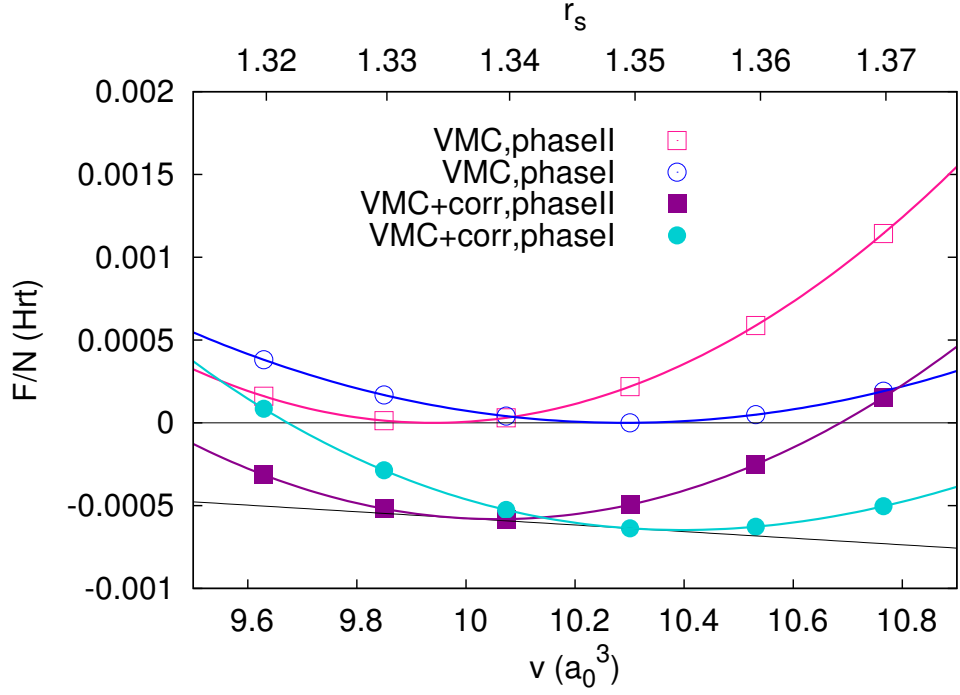}
\caption{Common tangent construction applied to the specific
Helmholtz free energy curves for the system with  VMC energies
(open symbols) and to the corrected curves (closed symbols). To
enhance the curvature a linear term, $\ell(v)=0.00934v -
0.09303$, has been subtracted to all curves. The common
tangents are represented by the black lines.  After the
addition of the correction the coexistence pressure changes
from $P^{c}=275.5(5)GPa$ to $P^{c}=281(1)GPa$. The qualitative
coexistence pressure inferred from the RQMC data is
$P^{c}\simeq 280(1)GPa$.} \label{fig:maxwell}
\end{figure}

\begin{figure}
\center
\includegraphics[width=0.49\textwidth]{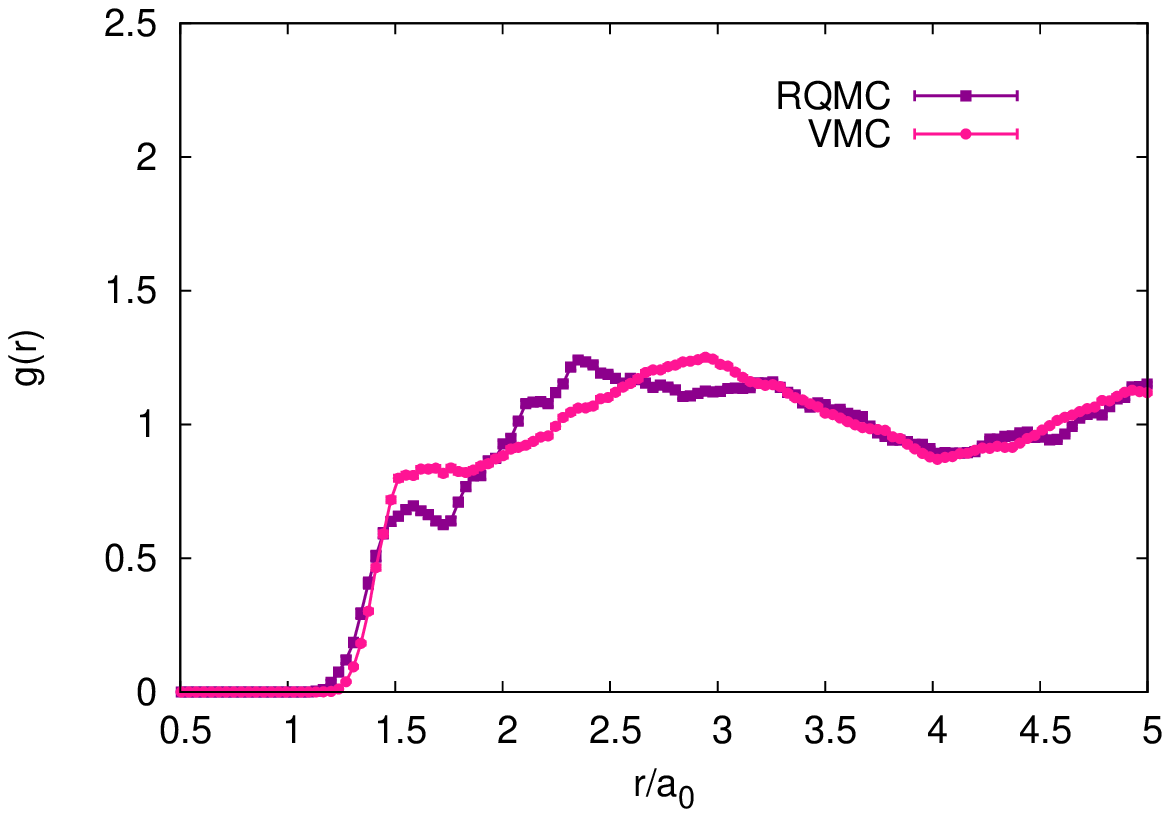}
\includegraphics[width=0.49\textwidth]{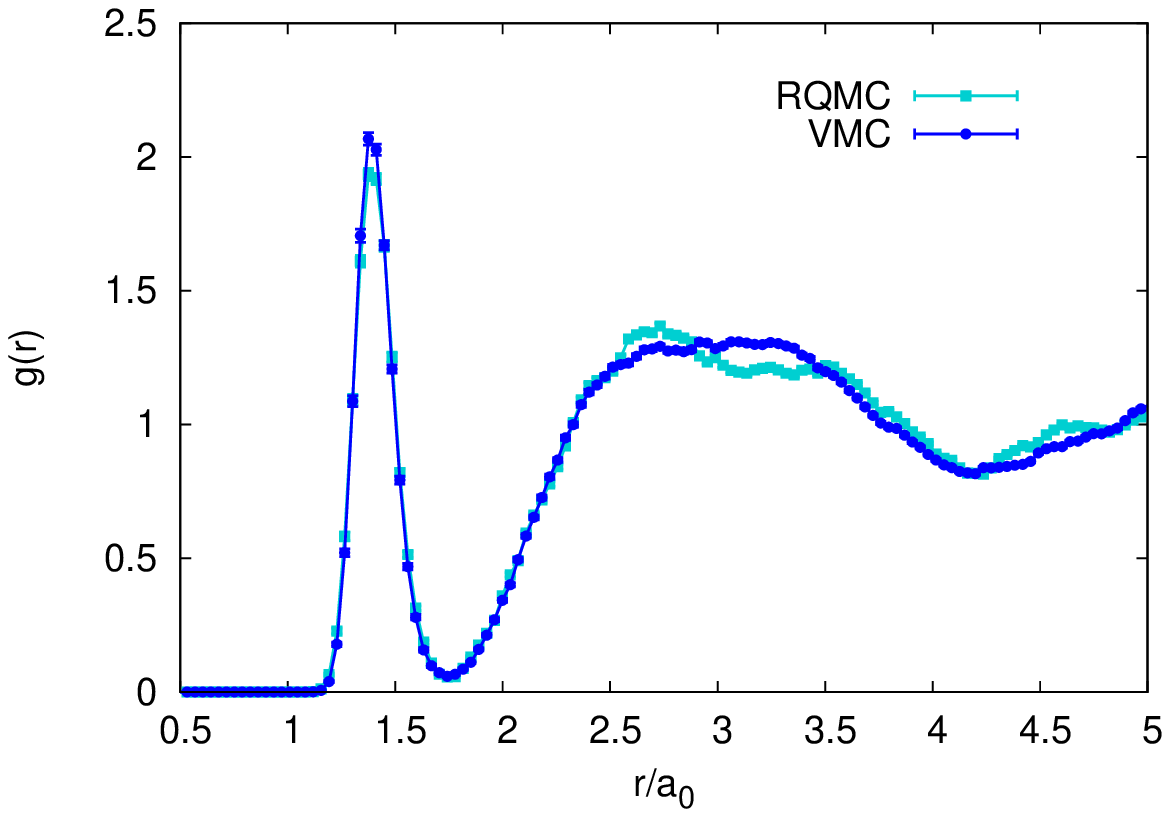}
\caption{Comparison between the proton-proton radial
distribution functions $g(r)$  obtained with the two electronic
methods at $r_{s}=1.32$ (metallic phase (II), left panel) and
$r_{s}=1.39$ (molecular insulating phase (I), right panel).}
\label{fig:grcfr}
\end{figure}

\begin{figure}[h]
\center
\includegraphics[width=0.9\textwidth]{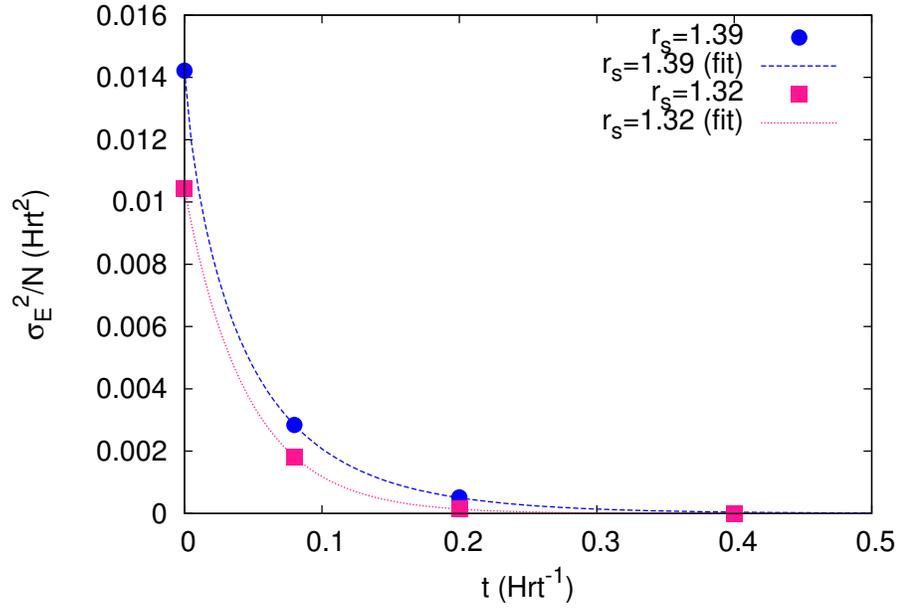}
\caption{Average variance of the BO energy per particle $\sigma_E^{2}/N$ as a function of the total projection time $t$, at $T=600K$ and for two densities at the two edges of the liquid-liquid transition: $r_{s}=1.39$ (blue circles, molecular insulating phase I) and $r_{s}=1.32$ (pink squares,  metallic phase II). The behavior of $\sigma_{E}^{2}/N$ with $t$ can be fitted  as a stretched exponential $s(t)=ae^{-bt^{c}}$, with $a=0.01422H^{2}$, $b=12.05H$, $c=0.797$ at $r_{s}=1.39$ and $a=0.01042H^{2}$, $b=21.1H$, $c=0.987$ at $r_{s}=1.32$. Values at finite $t$ are corrected for the nonzero time step error.}
\label{fig:sigmas}
\end{figure}

To test the ability of the correction formula \eqref{eq:rqmcFE}
to predict the RQMC coexistence pressure of the metallization
transition at $T=600K$ we proceed as follows.
\begin{itemize}
\item[a)]  We assume that the coexistence value for the pressure and the specific volumes from the
qualitative visual analysis above are correct and provide our reference results for the VMC system.
With this assumption and using the fits to the EOS's above, we reconstruct the Helmholtz free energy curves
for the two phases up to a constant value, by integrating the VMC-EOS's.
For a given phase $\alpha=I,II$ the free energy per particle at the specific volumes $v$ is related to the free
energy per particle at coexistence volume $v^c_{\alpha}$ by
\beq
f_{\alpha}(v)-f_{\alpha}(v^c_{\alpha})&=&-\int_{v^c_{\alpha}}^{v}  P_{\alpha}(\nu)d\nu  \label{eq:dfa0}  \nonumber\\
&=& - p_{\alpha}(v-v^c_{\alpha})-\frac{q_{\alpha}}{2}\left(v^{2}-(v^c_{\alpha})^{2}\right)-\frac{r_{\alpha}}{3}\left(v^{3}-(v^c_{\alpha})^{3}\right)
\label{eq:dfa}
\eeq
where the final expression has been obtained by integrating a second order polynomial fit of the equation of state.
The free energies of two coexisting phases can be referred to a single reference value noticing that at coexistence
\beq
f_I(v^c_I)-f_{II}(v^c_{II})=P^c(v^c_{II}-v^c_I)
\eeq
The resulting curves are plotted in Figure \ref{fig:maxwell}.
The magenta open squares represent the free energy per particle
of phase II and the blue open circles the free energy per
particle of phase I. In order to enhance their curvature and
make the visualization of the common tangent easier,we
subtracted  the same linear common tangent
$\ell(v)=P^{c}(v-v_{I}^{c})=0.00934H/a_{0}^{3} \ v -0.09303H$
from each curves. This common tangent (represented by the
horizontal black line in Figure \ref{fig:maxwell}) is, by
construction, the coexistence pressure corresponding to the
plateau in  the VMC pressure curve as discussed above.

\item[b)]  Once the free energy curves for the VMC system are known we can apply the correction term in eq. (\ref{eq:rqmcFE})
to obtain the RQMC free energy curves. The most convenient strategy is to compute the correction at a single fixed specific volume
for each phase $\tilde{v}_{\alpha}$ and to reconstruct the RQMC-free energy curves by integrating the RQMC-EOS of the two phases.
To this aim, at a single fixed density in each phase we perform several RQMC calculations for increasing projection time to numerically compute the correction term
\beq
\delta_{\alpha}(\tilde{v}_{\alpha})=f_{\alpha}^{RQMC}(\tilde{v}_{\alpha})-f_{\alpha}^{VMC}(\tilde{v}_{\alpha})=-\frac{1}{N}\lim_{t\to\infty}\int_0^t ds \overline{\left(\sigma^2_E(s,\tilde{v}_{\alpha})\right)_s}
\eeq
The values of the reference volumes $\tilde{v}_{\alpha}$ must
be taken far enough from the coexistence region to avoid
crossing the phase boundary during the time integration
process. To satisfy this requirement we have chosen
$\tilde{v}_{II}= 9.634 a_{0}^{3} \ (r_{s}=1.32)$ for the
metallic liquid (phase II) and $\tilde{v}_{I}=11.249 a_{0}^{3}
\ (r_{s}=1.39)$ for the insulating liquid (phase I). In Figure
\ref{fig:grcfr} we compare the proton-proton radial
distribution functions $g(r)$ at the two densities as obtained
by CEIMC with VMC and RQMC electronic energies.  We can see
that at both densities the system is in the same phase
regardless of the electronic method used.  At
$\tilde{v}_{I}=11.249 a_{0}^{3}$ both $g(r)$ present a sharp
peak centered at $r\sim 1.4a_{0}$, which indicates that the
system is in the molecular phase; at $\tilde{v}_{II}= 9.634
a_{0}^{3} $ the molecular peak has completely disappeared and
the form of $g(r)$ is typical of a monoatomic fluid.

In order to evaluate the integral in  eq. \eqref{eq:rqmcFE}, at
each reference specific volume we performed additional
CEIMC-RQMC simulations with projection times
$t=0.08,0.20,0.40H^{-1}$ and a time step $\tau=0.02H^{-1}$.
The behavior of $\sigma_{E}^{2}/N$ with $t$, corrected by
finite time step error, is shown in Figure \ref{fig:sigmas} and
is well fitted by a stretched exponential function:
$s(t)=ae^{-(t/b)^{c}}$ and its time integral is obtained by
using the relation \beq \int_{0}^{\infty} e^{-(t/b)^{c}} dt =
\frac{b}{c}\Gamma\left(\frac{1}{c}\right) \eeq where
$\Gamma(x)$ is the Gamma function.  The values of the fitting
parameters in the two phases are $a_I=0.01422H^{2}$,
$b_I=0.0440H^{-1}$, $c_I=0.797$ at $v=\tilde{v}_{I}$ and
$a_{II}=0.01042H^{2}$, $b_{II}=0.0455H^{-1}$, $c_{II}=0.987$ at
$v=\tilde{v}_{II}$ and the RQMC-VMC Helmholtz free energy
difference per particle results to be
$\delta_{I}(\tilde{v}_{I})=- 0.71(1)mH$ for the molecular
insulating phase (phase I) and $\delta_{II}(\tilde{v}_{II})
=-0.48(1)mH$ for the metallic phase (phase II).

\item[c)] finally we use the RQMC-EOS to construct the RQMC free energy curves and to determine the coexistence
pressure and volumes between the phases I and II by the common tangent construction.
In figure \ref{fig:maxwell} the RQMC free energy curves of
phase I and II are represented by the closed turquoise circles
and the closed purple squares, respectively. Despite the fact
that the correction from the VMC free energy is quite
small\footnote{We can have an idea of the relative amount of
the free energy corrections by comparing the difference
$\delta_{I}(\tilde{v}_{I})-\delta_{II}(\tilde{v}_{II})=0.23(2)mH$
with the VMC free energy difference between phase I and II at
the same thermodynamic points,
$f_{I}(\tilde{v}_{I})-f_{II}(\tilde{v}_{II})\sim 14.5mH$: the
contribution of the correction is then quite small as expected,
about  $1.5\%$ of the VMC free energy differences. }, less than
$1mH$ per particle over the range of volumes under analysis, it
has a significant effect on the coexistence: the common tangent
construction applied to those curves provides a coexistence
pressure of $P^{c}=281(1)GPa$ and specific volumes of the two
phases at coexistence of $v_{I}^{c}=10.32(4)a_{0}^{3}$ and
$v_{II}^{c}=9.99(4)a_{0}^{3}$. Those values are in excellent
agreement with the values directly inferred from the RQMC
pressure curve ($P^{c}=280(1)GPa$,
$v_{I}^{c}=10.31(4)a_{0}^{3}$ and
$v_{II}^{c}=10.00(4)a_{0}^{3}$), being the differences between
the two predictions within error bars.
\end{itemize}

\section{Discussion and Conclusion}\label{sec:conclusions}
Within the Coupling Constant Integration strategy and for the
Coupled Electron-Ion Monte Carlo method, we have derived an
original relation to obtain the free energy of an ab-initio
system with Reptation Quantum Monte Carlo electronic energies
from the knowledge of the free energy of the same system but
with Variation Monte Carlo electronic energies, a considerably
less demanding calculation. This relation will be useful in
improving the accuracy of the CEIMC method in tracing
transition lines, therefore predicting phase diagrams, without
increasing too much the already large computational
requirements of the method. In order to test the use of the new
relation and to illustrate the procedure to correct transition
points we have considered the liquid-liquid transition in high
pressure hydrogen, a first order transition between a lower
pressure, molecular and insulating phase (I) and an higher
pressure, partially dissociated and conducting phase (II). We
have reported new CEIMC results along the isotherm at $T=600K$
in the density range corresponding to $r_s\in [1.31,1.40]$
across the transition, an isotherm not considered before. Not
presenting metastable states, this transition can be detected
by the appearance of a plateau in the pressure-volume curve and
the values for coexistence pressure and for the specific
volumes of the phases at coexistence can be inferred without
performing  free energy calculations. Nonetheless, as detailed
in the paper, it can still be used to illustrate our procedure
for correcting transition lines. In particular, we get
excellent agreement between the RQMC critical pressure inferred
from the data ($P^c=280(1)GPa$) and the value obtained by our
procedure ($P^c=281(1)GPa$).  A similar agreement is obtained
for the specific volumes of the two phases at coexistence.

The close agreement between VMC and RQMC based Equation of
States (EOSs) and transition points results from the high
quality of our trial wave function for high pressure hydrogen
even across the metal-insulator transition, a notoriously
difficult region for DFT based first principle methods. This
agreement strongly supports the use of VMC in CEIMC
investigations of the phase diagram of hydrogen at high
pressure, allowing calculations on larger systems or systems
with quantum protons within the Path Integral formalism
\cite{PierleoniLNP2006}. Moreover the formula derived in the
present work can be used to improve the CEIMC-VMC transition
lines to RQMC accuracy.

\section*{Acknowledgments}
CP is supported by the Italian Institute of Technology (IIT)
under the SEED project grant n 259 SIMBEDD - Advanced
Computational Methods for Biophysics, Drug Design and Energy
Research. DMC is supported by DOE grant DE-FG52-09NA29456.
This work was performed in part under the auspices of the US DOE by LLNL
under Contract DE-AC52-07NA27344.
Financial support from the Erasmus Mundus Program-Atosim is
acknowledged. Computer resources were provided by CASPUR
(Italy) within the Competitive HPC Initiative, grant number:
cmp09-837, and by the DEISA Consortium (www.deisa.eu),
co-funded through the EU FP6 project RI-031513 and the FP7
project RI-222919, through the DEISA Extreme Computing
Initiative (DECI 2009). \nocite{*}
\bibliography{ceimc}

\begin{table}[h]
\center \caption{Pressures along the $T=600K$ isotherm, for
RQMC (third column) and VMC (fourth column) data sets.  In the
first column, the values of $r_{s}$ and in the second the
corresponding specific volumes $v$ are reported.}
\label{tab:PPTdata}
\begin{tabular}{cccc}
\hline
\hline
$r_{s}$ & $v(a_{0}^{3})$ &  $P_{RQMC}(GPa)$ & $P_{VMC}(GPa)$\\
\hline
\hline
1.310 & 9.417  &331.6(6)& 327.2(6)\\
1.320  & 9.634  &311.4(4)& 305.2(6)\\
1.330 & 9.855 &293.9(6)&284.0(6) \\
1.340 & 10.08 &283.8(7)& 275.9(6)\\
1.350 & 10.31 &277.2(7)& 275.0(6)\\
1.355 & 10.42 &277.8(8) & -\\
1.360 & 10.54 &271.5(6)&262.8(6)\\
1.370 & 10.77 &253.1(6)&252.6(6) \\
1.380 & 11.01 &240.7(6)&239.2(5) \\
1.390 & 11.25 &231.0(6)&229.6(6) \\
1.400 & 11.49 &220.4(6)&217.8(6)\\
\hline
\hline
\end{tabular}
\end{table}
\newpage

\newpage

\newpage

\newpage
\newpage

\end{document}